\documentclass[preprint,showpacs,preprintnumbers,amsmath,amssymb,nofootinbib,showkeys,aps]{revtex4}

\usepackage{epsfig}
\usepackage{dcolumn}
\usepackage{url}
\usepackage{bm}
\usepackage{subfigure}

\numberwithin{equation}{section}		

\makeatletter       

\renewcommand{\p@subsection}{}

\renewcommand{\p@subsubsection}{}

\makeatother

\def\fsu{$\widetilde{SU}(5)$}

\def\warp{k \pi r_c}

\def\szorb{$S^1/\mathcal{Z}_2$}
\def\szzorb{$S^1/(\mathcal{Z}_2 \times \mathcal{Z}_2')$}

\def\beqn{\begin{eqnarray}}
\def\eeqn{\end{eqnarray}}

\begin{document}
\vskip 0.5truecm
\preprint{BU-HEPP 06-06}
\preprint{CASPER 06-02}
\vskip 0.5truecm
\title{Randall-Sundrum and Flipped SU(5)}
\vskip 0.5truecm
\author{Ben Dundee}
	\email{dundee.1@osu.edu}
\author{Gerald Cleaver}%
 \email{Gerald_Cleaver@baylor.edu}
\affiliation{
The Center for Astrophysics, Space Physics, and Engineering Research\\
One Bear Place $\#$97316\\
Department of Physics, Baylor University\\
Waco, TX 76798\\
\\
}


\vskip 0.5truecm

\begin{abstract}
In this letter, we construct a model based on a flipped $SU(5)$ partial grand unified theory, within the framework of the Randall-Sundrum (RS1) proposal.  Breaking of $\widetilde{SU}(5)$ is achieved using a bulk scalar field in the \textbf{10} of $SU(5)$, $\Phi$, which gains a vacuum expectation value $\left\langle \Phi\right\rangle \sim 3 \times 10^{15}$ GeV.  We are able to retain the successes of the \fsu~phenomenology, namely the elimination of the doublet-triplet splitting problem and the confinement of all fields to the smallest (\textbf{1}, $\mathbf{\overline{5}}$, and \textbf{10}) representations of $SU(5)$.  We derive the beta functions, and point out some constraints on bulk matter content implied by the runnings (and positivity) of the five dimensional coupling.  Finally, we comment on baryon decay and show the fine-tuning problem required to prevent an exponentially short proton lifetime.

\end{abstract}


\maketitle

\tableofcontents

\section{\label{sec:intro}Introduction}
The Randall-Sundrum (RS1) proposal \cite{Randall:1999ee} represents a beautiful geometrical solution to the hierarchy problem of particle physics.  Specifically, by embedding a four dimensional Minkowski space in a higher dimensional anti-de Sitter ($AdS$) bulk, one is able to suppress the weak scale by a factor of $e^{-k \pi r_c}$, with $k$ and $r_c$ the warp factor and compactification radius, respectively.  The normal argument goes like this: the bulk metric is given by
\begin{eqnarray}
	ds^2 = e^{-2 k z} \eta_{\mu \nu} dx^{\mu} dx^{\nu} + dz^2,
	\label{linesegment}
\end{eqnarray}
where the fifth dimension ($z$) is an orbifolded circle (\szorb).  If the spectrum contains a bulk scalar, $H$, its action looks like this:
\begin{eqnarray}
	\mathcal{S} \supset \int d^4x \int^{\pi r_c}_{0} dz \sqrt{-G} \left(G^{MN}\partial_{M} H^{\dagger} \partial_{N} H + m^2 H^{\dagger} H\right),
	\label{higgsaction1}
\end{eqnarray}
where $G_{MN}$ is the full 5-dimensional metric and is given by
\begin{subequations}
\label{gmnandinverse}
	\begin{eqnarray}
		G_{MN} = e^{-2 k z} \rm{diag}(-1,+1,+1,+1,e^{2 k z})\;, \label{gmn} \\
		G^{MN} = e^{2 k z} \rm{diag}(-1,+1,+1,+1,e^{-2 k z}). \label{invgmn}
	\end{eqnarray}
\end{subequations}
If our universe exists in the IR limit of (\ref{higgsaction1}), then we evaluate the action on the IR brane, namely where $z = \pi r_c$, and find
\begin{eqnarray}
	\mathcal{S}_{eff} \sim \int d^4x \left(e^{-2 k \pi r_c} 	\partial^{\mu}H^{\dagger}\partial_{\mu}H + e^{-4 k \pi r_c} m^2 H^{\dagger}H\right).
	\label{leeftforhiggs}
\end{eqnarray}
Redefining the scalar field $H \rightarrow e^{-k \pi r_c} H$ we see
\begin{eqnarray}
	\mathcal{S}_{eff} \sim \int d^4x \left(\partial^{\mu}H^{\dagger}\partial_{\mu}H + e^{-2 k \pi r_c} m^2 H^{\dagger}H\right).
	\label{higgsaction2}
\end{eqnarray}
The mass (which is on the order of the Plank mass, as we expect from dimensional analysis) is now weighted by the warp factor on the IR brane where we would observe it, $m \rightarrow m e^{-k \pi r_c}$, and can be tuned to give phenomenologically acceptable values.  The stabilization of the brane separation has already been addressed in \cite{Goldberger:1999uk}, so that the choice of 
\begin{eqnarray}
	k \pi r_c = \log \left[\frac{M_{UV}}{M_{IR}}\right] \sim 11 \pi
	\label{GWMech}
\end{eqnarray}
is well-motivated.  
\subsection{\label{sec:matterandforces}Adding Matter and Forces}
One can add spin $\frac{1}{2}$ fermions \cite{Grossman:1999ra} and spin 1 bosons \cite{Davoudiasl:1999tf,Pomarol:1999ad} to the spectrum in much the same way.  Upon doing so, we find that the spin 1 bosons have flat wavefunctions in the fifth dimension.  A spin $\frac{1}{2}$ fermion whose wavefunction is symmetric about the \szorb~orbifold symmetry has the general form:
\begin{eqnarray}
	\Psi_5 \sim e^{(\frac{1}{2} - c)k z} \psi_4.
	\label{fermwf}
\end{eqnarray}
Here, $\Psi_5 \left(\psi_4\right)$ is the five- (four-) dimensional wavefunction.  The $c-$values ($c\in\left[0,1\right]$) dictate about which brane the wavefunction is localized, and thus the observed mass in the low energy effective field theory.\footnote{The boundary mass terms in the 5-d action are defined as $m=ck$.}  By picking numbers of $\mathcal{O}(\frac{1}{2})$, one is able to generate mass hierarchies which are put into the Standard Model (SM) by hand.
\subsection{\label{sec:uni}SU(5) Unification}
It is well known that the SM matter fits into the \textbf{1}, the $\overline{\mathbf{5}}$ ($\overline{\chi}$) and a \textbf{10} ($\psi$) reps of $SU(5$) \cite{Georgi:1974sy}.
\begin{equation}
	\begin{array}{ccc}
		\bar{\chi} =	\left(\begin{array}{c} 
					 	\bar{d}^1\\
					 	\bar{d}^2\\
					 	\bar{d}^3\\
					  e\\
					 	\nu_e \end{array}\right)_L
		&\psi = \left(\begin{array}{ccccc}
						0&\bar{u}_3&-\bar{u}_2&-u_1&-d_1\\
						-\bar{u}_3&0&\bar{u}_1&-u_2&-d_2\\
						\bar{u}_2&-\bar{u}_1&0&-u_3&-d_3\\
						u_1&u_2&u_3&0&-\bar{e}\\
						d_1&d_2&d_3&\bar{e}&0\end{array}\right)_L
		&\bar{\nu_e} = \mathbf{1}
		\end{array}
	\label{su5reps}
\end{equation}
Traditionally, the breaking of $SU\left(5\right)$ is achieved when some scalar field takes on a vev.  In the simplest example, a scalar $\Sigma$, transforming in the \textbf{24} (adjoint) of $SU\left(5\right)$, takes on a vev of the form
\begin{eqnarray}
	\left\langle \Sigma\right\rangle =  v \left(\begin{array}{ccccc}
																			-\frac{2}{3}&0&0&0&0\\
																			0&-\frac{2}{3}&0&0&0\\
																			0&0&-\frac{2}{3}&0&0\\
																			0&0&0&1&0\\
																			0&0&0&0&1\end{array}\right),
	\label{sigmavev}
\end{eqnarray}
by minimizing an assumed potential \cite{Mohapatra:1986uf}.
This form of the vev for the scalar $\Sigma$ is justified \textsl{ex post facto}---it leaves the gluons, the $W$'s and the $B$ all massless and preserves the SM at energies $<< M_{GUT}$, while giving the other 12 generators of $SU\left(5\right)$ (the $X$'s and $Y$'s) masses $\sim \frac{25}{9} v^2$.  Finally, a higgs in the fundamental ($\textbf{5}$) representation achieves electroweak symmetry breaking, giving the $W^{\pm}$ and the $Z^0$ bosons mass $\sim$ 100 GeV, fourteen orders of magnitude smaller than the masses of the $X$ and $Y$ bosons.  
\subsection{\label{sec:breakingsu5}RS GUTS and Breaking SU(5) with Boundary Conditions}
The question of unification in RS1 models has been addressed in several places, and in several different incarnations.  The first examples of unification in RS1 were based on $SU\left(5\right)$ GUTs \cite{Agashe:2002pr}.  Supersymmetric $SU\left(5\right)$ unification has been investigated \cite{Goldberger:2002pc} and more recently, some detailed investigations of $SO\left(10\right)$ GUTs were preformed \cite{Agashe:2004bm}.  Generally, the breaking of the GUT symmetry down to the SM has been achieved by assigning different boundary conditions to the fields appearing in the representation.

In typical RS1 GUT constructions, fermion fields are allowed to live in the bulk.  One typically replaces the \szorb~orbifold with an \szzorb~ orbifold.  The fields are assigned von Neumann (+) or Dirichlet (--) boundary conditions about the branes at the ends of the $AdS$ space.\footnote{Equivalently, we could say ``...parities under the $S^1/(\mathcal{Z}_2 \times \mathcal{Z}_2')$~orbifold..." \cite{Agashe:2002pr}.}  These boundary conditions are specified in doublets, with the first entry corresponding to the UV brane, and the second corresponding to the IR brane.  Physical (massless) modes have (+ +) boundary conditions.  Other sets of boundary conditions lead to Planck scale massive fields, which do not contribute to the low energy phenomenology.  Also, the individual fields in each representation are allowed to take on different boundary conditions.

Breaking the GUT symmetry with boundary conditions requires that one add copies of representations to the spectrum, and the quarks and leptons (from the \textit{same} generation) in the SM come from \textit{different} copies of identical representations of the underlying GUT symmetry group.  For example, one of the models in \cite{Agashe:2002pr} has
\begin{subequations}
\label{ADS1}
	\begin{eqnarray}
		\overline{\mathbf{5}}_1 = L^{++}_1 + d^{+-}_1 \label{ADS1.1},\\
		\overline{\mathbf{5}}_2 = L^{+-}_1 + d^{++}_1 \label{ADS1.2}.
	\end{eqnarray}
\end{subequations}
The SM states have (+ +) boundary conditions while the other fields have (+ --) boundary conditions.  While all of these models elegantly incorporate the features of the RS1 proposal, they all suffer from this seemingly universal problem of representation proliferation.  For example, the $SO\left(10\right)$ model in ref. \cite{Agashe:2004bm} required 6 copies of each \textbf{16} for each generation!

This proliferation of representations does solve some problems.  Experimentalists have given us strict bounds on the proton's lifetime, $\tau_p > 6.7 \times 10^{33}$ years \cite{Hayato:1999az}, and one must be wary of higher dimensional operators in the effective field theory that violate these bounds.  For example, in 5-d one could write the following operator down (from $\overline{\textbf{5}} \times \textbf{5} \times \overline{\textbf{10}} \times \textbf{10} \supset \textbf{1}$):
\begin{eqnarray}
	\int d^4x dz \sqrt{-G} \frac{\overline{\Psi}_{\overline{5}}\Psi_5\overline{\Psi}_{\overline{10}}\Psi_{10}}{M^{3}_{5}}.
	\label{baryondecay}
\end{eqnarray}
We can evaluate this integral on the IR brane, using (\ref{fermwf}),
\begin{eqnarray}
	e^{\pi k r_c (4-c_1-c_2-c_3-c_4)} \int d^4x \frac{\overline{\psi}_{\overline{5}} \psi_{5}\overline{\psi}_{\overline{10}}\psi_{10}}{M^{2}_{Pl}}.
	\label{bdopinleeft}
\end{eqnarray}
The $c-$values are less than 1, making the coupling constant in the effective field theory exponentially large, and proton lifetime exponentially short.  By requiring quarks and leptons to come from different representations, there exist no physical (i.e. on the IR brane) quark-lepton mixing, as is typical of traditional GUTs.  It seems that there are no conventional baryon decay modes in this type of model, so the prediction is that experiments like Super-Kamiokande \cite{Hayato:1999az} will \textit{never} see $p \rightarrow$ leptons + mesons.  If it did, then one would either have to find some other way to suppress these processes, or accept the exponentially tuned Yukawa couplings required by terms like (\ref{bdopinleeft}), and abandon the RS1 paradigm of no fine-tunings.

Finally, it should be noted that there have been investigations into breaking the GUT by turning on the vev of a bulk scalar field \cite{Agashe:2002pr, Contino:2002kc}.
\subsection{\label{sec:fsu5}A Review of Flipped SU(5)}
Flipped SU(5) (or $\widetilde{SU}(5)$) \cite{Barr:1981qv, Derendinger:1983aj} unification has been studied extensively in the context of string model building \cite{Antoniadis:1987tv,Antoniadis:1988tt,Antoniadis:1989zy,Leontaris:1990bw,Bailin:1990qb,Burwick:1990vy,Rizos:1990xn,Lopez:1991ac,Lopez:1991ec,Antoniadis:1991fc,Lopez:1992kg,Hatzinikitas:1992ip,Lopez:1995pr,Lopez:1997hq,
Ellis:1997ni,Ellis:1999ce,Cleaver:2000sc,Faraggi:2002ah,Faraggi:2002dg,Chen:2005ab,Ellis:2004cj,Chen:2005mm,Chen:2005cf,Chen:2006ip,Huang:2006nu,Shafi:2006dm,Cvetic:2006by,Kim:2006hv,Kim:2006hw,Kim:2006hv,Blumenhagen:2006ux}.\footnote{The first realistic examples of the string derived standard model \cite{Cleaver:1998sa} also came from $SO(10)$ embeddings of $\widetilde{SU}(5)$, obtained from the free fermionic heterotic string \cite{Kawai:1986ah,Antoniadis:1986rn,Kawai:1987ew,Antoniadis:1987wp}.}  The fermions still fit into the \textbf{1}, $\overline{\textbf{5}}$ and \textbf{10} reps of $SU\left(5\right)$.  The only difference is that the anti-neutrino and the electron, and the up- and down-type quarks exchange places in their respective reps.  So, we have
\begin{equation}
	\begin{array}{ccc}
		\overline{\chi} =	\left(\begin{array}{c} 
					 	\bar{u}^1\\
					 	\bar{u}^2\\
					 	\bar{u}^3\\
					  \nu_e\\
					  e\\
					 	\end{array}\right)_L,
		&\psi = \left(\begin{array}{ccccc}
						0&\overline{d}_3&-\overline{d}_2&-d_1&-u_1\\
						-\overline{d}_3&0&\overline{d}_1&-d_2&-u_2\\
						\overline{d}_2&-\overline{d}_1&0&-d_3&-u_3\\
						d_1&d_2&d_3&0&-\nu_e\\
						u_1&u_2&u_3&\nu_e&0\end{array}\right)_L,
		&\bar{e} = \mathbf{1}
		\end{array}
	\label{fsu5reps}
\end{equation}
Two things are important here: First, it is immediately obvious that the (electric) charges of the fields in the reps (\textit{cf} the $\overline{\textbf{5}}$) do not trace to zero.  This means that we must add another generator (proportional to the identity) to ensure freedom from anomalies.  The actual gauge group of this model is
\begin{eqnarray}
	SU(5) \times U(1)_{\tilde{Y}}
\end{eqnarray}
Second, because there is now a (color and electric) neutral member in the \textbf{10}, we may use this non-adjoint rep for higgsing the GUT.  Historically, this represents the first time a representation with a dimension less than that of the adjoint was used for symmetry breaking in a grand unified theory.  Other than these two minor differences, much of the phenomenology, as well as many of the predictions of traditional $SU\left(5\right)$ GUTs, are preserved in $\widetilde{SU}(5)$.

Because we will need the covariant derivative for calculating the $\beta$ functions of our model,\footnote{It is quite difficult to find the proper form of this in the literature, so we have also listed it here for posterity's sake!} we list it here for the \textbf{10}:
\begin{eqnarray}
	D_{\mu} \Phi = \partial_{\mu} \Phi - i g \left\{A_{a\mu}\frac{\lambda_a}{2} \Phi + \Phi A_{a\mu}\frac{\lambda_a^T}{2}\right\} - i \tilde{g} \tilde{Y} \tilde{B}_{\mu} \Phi.
	\label{fsu5covderiv10}
\end{eqnarray}
The $\lambda^a$'s are the 24 generators of SU(5), and the $A^a_{\mu}$'s are the corresponding gauge bosons.  The extra $U(1)_{\tilde{Y}}$'s gauge boson is denoted by $\tilde{B}_{\mu}$.  As a quick aside---the higgsing of \fsu~ to the SM can be worked out using this form of the covariant derivative, and the form of the higgs vev:
\begin{eqnarray}
	\left\langle \Sigma\right\rangle =  \left(\begin{array}{ccccc}
																			0&0&0&0&0\\
																			0&0&0&0&0\\
																			0&0&0&0&0\\
																			0&0&0&0&v\\
																			0&0&0&-v&0\end{array}\right).
\end{eqnarray}

SUSY $\widetilde{SU}(5)$ unification in extra dimensions has been studied in \cite{Barr:2002fb} and \cite{Dorsner:2003yg}, but this analysis was in the presence of a flat extra dimension, on an orbifolded circle ($S^1 / \mathcal{Z}_2 \times \mathcal{Z}_2'$).  The main difference between these two approaches is the form of the Kaluza-Klein modes and their contributions to the $\beta$ funcions (see \cite{Agashe:2002bx} and \cite{Choi:2002ps} for more details)---in the RS1 proposal the masses of the KK modes are solutions to combinations of Bessel functions, whereas in the flat case
\begin{eqnarray}
	m^2_n = m^2 + \frac{n^2}{R^2},
	\label{flatkkmasses}
\end{eqnarray}  
for $n = 0,1,2,...$.  Generally, one would like to unify the $\widetilde{SU}(5)$ theory in some higher dimensional GUT, like $SO\left(10\right)$.  Here we will only work at the partial GUT level, and leave the problem of $SO\left(10\right)$ embedding in the RS1 context to a future study.

\section{\label{sec:model}The Model}
The approach we take in this study is one of minimalism.  We choose the phenomenologically well-motivated $\widetilde{SU}(5)$ partial GUT as a starting point, and achieve breaking with a higgs field $\Phi$ in the \textbf{10}. In Section \ref{subsec:spectrum} we analyze the spectrum of our model.  In Section \ref{subsec:gcr} we compute the beta function of our model, using the results of \cite{Contino:2002kc, Agashe:2002bx, Goldberger:2002hb, Choi:2002ps}.  We show that the beta functions imply some constraints on the bulk matter content of RS1 GUTs in general (whether the breaking be due to boundary conditions or a GUT scalar), which we comment on in Section \ref{subsec:constraints}---to our knowledge, these constraints have not been pointed out in the literature.  Finally, we show that (as expected) the proton is much too short-lived for this model to be realistic, in Section \ref{subsec:numassandpdecay}.

\subsection{\label{subsec:spectrum}The Spectrum}
In our model, we consider a \fsu~partial GUT living in the background of Equation (\ref{linesegment}).  The breaking of \fsu~is accomplished with a bulk scalar field, called $\Phi$, that takes on a vev at some intermediate scale $M_*$.  Note that, in general, $M_*$ is less than the GUT scale, $M_{GUT}$, but it is not completely unreasonable that one push $M_*$ up to $M_{GUT}$---this would eliminate the embedding of \fsu~into some larger symmetry, like $SO\left(10\right)$.  We take a minimal matter content, as in Equation (\ref{fsu5reps}).

\subsection{Gauge Coupling Renormalization}
\label{subsec:gcr}
Calculating the one-loop corrections to the vacuum-to-vacuum polarizations is relatively straightforward, and has been done for the (extremely popular) case of scalar QED in several places \cite{Agashe:2002bx, Choi:2002ps, Contino:2002kc, Goldberger:2002hb}.  The difference between the standard QFT calculation and the RS1 calculation is the appearance of a tower of KK modes.  For fields that are even about the orbifold $\mathcal{Z}_2 \times \mathcal{Z}'_2$~symmetry, the KK masses are solutions to \cite{Gherghetta:2000qt}
\begin{eqnarray}
	b_{\alpha} \left(m_n\right)= b_{\alpha} \left(m_n e^{\pi k r_c}\right),
	\label{KKMass}
\end{eqnarray}
where
\begin{eqnarray}
	b_{\alpha} \left(m_n\right) = -\frac{\left(-r + \frac{s}{2}\right) J_{\alpha} \left(\frac{m_n}{k}\right) + \frac{m_n}{k} J_{\alpha}' \left(\frac{m_n}{k}\right)}{\left(-r + \frac{s}{2}\right) Y_{\alpha} \left(\frac{m_n}{k}\right) + \frac{m_n}{k} Y_{\alpha}' \left(\frac{m_n}{k}\right)},
	\label{balpha}
\end{eqnarray}
and the constants are given by
\begin{eqnarray}
	\begin{array}{ccc}
		\alpha = \left\{ \begin{array}{cc}
											\sqrt{4 + a}&\rm{spin}~0\\
											\big|c \pm \frac{1}{2}\big|&\rm{spin}~\frac{1}{2}\\
											\sqrt{1 + d}&\rm{spin}~1\end{array} \right.,&
		r = \left\{ \begin{array}{cc}
											b&\rm{spin}~0\\
											\mp c&\rm{spin}~\frac{1}{2}\\
											0&\rm{spin~}1\end{array}\right. ,&
		s = \left\{ \begin{array}{cc}
											4&\rm{spin}~0\\
											1&\rm{spin}~\frac{1}{2}\\
											2&\rm{spin}~1\end{array}\right..
	\end{array}											
	\label{KKMD}
\end{eqnarray}
In this study, we are concerned with fields which have (+ +) boundary conditions on the $\mathcal{Z}_2 \times \mathcal{Z}'_2$ orbifold symmetry---there are similar expressions for the odd case.  The constants $a$, $b$, $c$ and $d$ come from the wavefunction's ``boundary mass'', and parameterize the field's profile in the fifth dimension---for example, we have already seen how $c$ is defined in Equation (\ref{fermwf}).  The constant $a$ for a scalar field is given by $a \equiv \frac{m^2_{\Phi}}{k^2}$, where $m_{\Phi}$ is the five dimensional scalar mass.  The constant $b = 2 + \alpha$, and $d \equiv \frac{M^2}{k^2}$, where $M$ is the gauge boson's mass---possibly zero.  Finally, $n = 1, 2, \ldots$.

\begin{figure}[t!]
	\centering
	\includegraphics{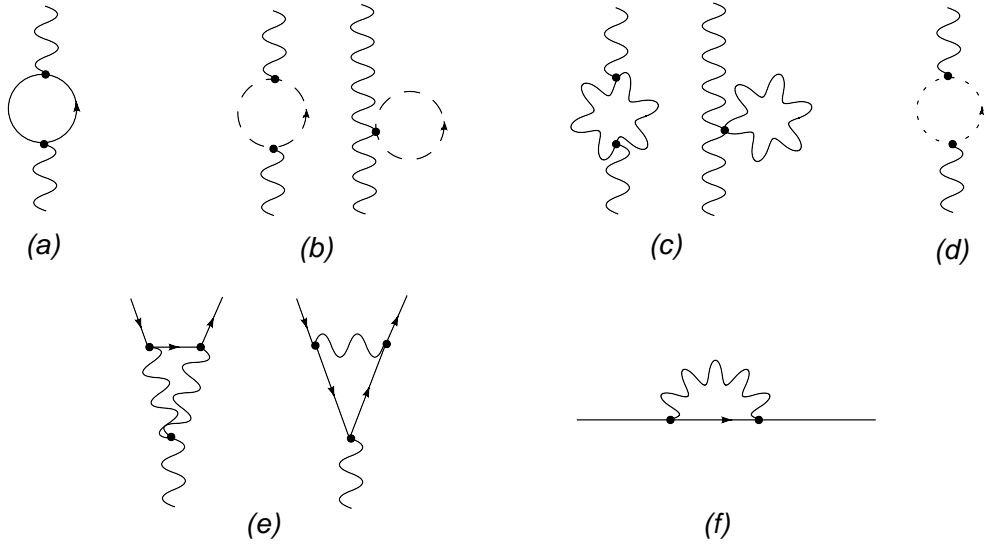}
	\caption{Possible graphs for calculating the massive contribution to the runnings of the couplings, $\alpha_i^{-1}$.  We must include contributions to the vector boson self-energy from fermions (a), scalars (b), \fsu~gauge bosons (c), and ghosts (d).  We also get contributions from the fermion-fermion-boson counterterm (e) and the fermion self-energy (f).  Note that we must include the KK mode sum in each of the loops.  Note that fields which lie in complete representations of the GUT do not contribute to the differential runnings of the couplings. (These graphs were generated using \cite{jaxodraw}.)}
	\label{fig:oneloop}
\end{figure}

Before we begin, it can be shown \cite{Pomarol:2000hp} that the leading contribution to the running of the couplings is logarithmic, as we find in the standard 4-d case---the corrections should contain terms $\sim \log\left[\frac{p}{\Lambda}\right]$, as long as we consider the regime where $p << \Lambda \lesssim k$, with $p$ some intermediate energy scale.  The general form of the running coupling constants in an RS1 background is given by \cite{Agashe:2002bx, Contino:2002kc, Choi:2002ps}:
\begin{eqnarray}
	\frac{1}{g^2_i\left(p^2\right)} = \frac{1}{k g_5^2} + \Delta_{UV} + \Delta_{IR} + \frac{1}{8 \pi^2} \left\{\Delta_{\rm{1-loop}} + \frac{\gamma_i}{24 \pi^3} \Lambda \pi r_c + b_i \log \left[\frac{\Lambda}{p}\right]\right\}.
	\label{genformrunning}
\end{eqnarray}
The $g_5$ is the bulk coupling constant, the $\Delta_{UV}$ and $\Delta_{IR}$ come from the couplings of the Maxwell tensors localized on the branes, and the $\Delta_{\rm{1-loop}}$ are the one-loop contributions from the graphs in Figure {\ref{fig:oneloop}--they arise because of the presence of Kaluza-Klein modes in the spectrum.  The linear divergences are regularization scheme dependent, and cannot be calculated within our effective field theory.  They are of $\mathcal{O}(\frac{M_*}{k})$ and will be ignored in what follows.\footnote{Equivalently, one could include these contributions in the redefinition (\ref{choiestimate}), as in \cite{Choi:2002ps}.}  The last term is the familiar non-Abelian beta function contributions.

In general, $\Delta_{UV}$, $\Delta_{IR}$ and $g_5$ are incalculable---they depend on some completion of the theory (possibly string theory).  We will take the incalculable parts of Equation (\ref{genformrunning}) to be \cite{Choi:2002ps}
\begin{eqnarray}
	\frac{1}{k g_5^2} + \Delta_{UV} + \Delta_{IR} \cong \frac{1}{g_{\widetilde{SU}(5)}^2} + \mathcal{O}\left(\frac{1}{8 \pi^2}\right).
	\label{choiestimate}
\end{eqnarray}

Now, let's compute the contribution of a massive scalar and its KK modes.  The relevant graphs are shown in Figure \ref{fig:oneloop} (b).  The Feynman rules for a scalar transforming in an arbitrary representation of a non-Abelian symmetry are just a straightforward modification of the rules for scalar QED.  We take our bulk scalar to have (+ +) boundary conditions.  If we compute the amplitude of the graphs in Figure \ref{fig:oneloop} (b), we find that the one loop correction, $\Delta_{\rm{1-loop}}^{\rm{scalar}}\left(q^2\right)$, for a massive scalar and its KK excitations is given by:\footnote{As was mentioned above, if the only field in the theory is the zero mode ($n=0$) scalar, then the sum consists of just one term---reducing the integral then gives the standard result.}
\begin{eqnarray}
	\Delta_{\rm{1-loop}}^{\rm{scalar}}\left(q^2\right) = \frac{g^2 C\left(r\right)}{\left(4 \pi\right)^{\frac{D}{2}}} \Gamma\left(2 - \frac{D}{2}\right) \int^{1}_{0} dx \sum_{\rm{KK~modes}} \left(\frac{\mu^2}{K_n}\right)^{2 - \frac{D}{2}} \left(1 - 2 x\right)^2,
	\label{PIQ2}
\end{eqnarray}
where
\begin{eqnarray}
	K_n = m^2_n + x \left(1-x\right)\left(-q^2\right) \equiv m^2_n + \chi^2.
	\label{delta}
\end{eqnarray}
$C(r)$ is the Dynkin index of the representation of the scalar field.  Sums of this form have been evaluated in \cite{Contino:2002kc, Goldberger:2002hb}, and Equation (\ref{PIQ2}) can be shown to be equal to
\begin{eqnarray}
	\nonumber
	\sum_{n} K_n^{\frac{D}{2} - 2} = \frac{1}{2} &+& \left(\frac{D}{2} - 2\right) \left[\log \left[f\left(i \chi\right)\right] + \log \left[\chi \pi \frac{e^{\frac{k \pi r_c}{2}}}{k}\right] + \log \chi \right]\\
	&+& \mathcal{O}\left(D - 4\right)^2,
	\label{closedsumofkkmodes}
\end{eqnarray}
where 
\begin{eqnarray}
	\chi = \sqrt{x \left(1 - x\right)^2\left(-q^2\right)}.
\end{eqnarray}
For a 5-d bulk scalar with (+ +) boundary conditions whose zero mode has mass $m_{\Phi}$, we have
\begin{eqnarray} 
	f \left(m_n\right) = \frac{1}{\pi \alpha} \left(\frac{e^{\warp}}{k^2}\right)^{\alpha - 1} \left(\frac{m_{\Phi}^2}{m_n^2} + \frac{2 + \alpha}{2 \alpha - 2}\right) .
	\label{beautifulf}
\end{eqnarray}
Because we are interested in the effect on the low energy effective field theory ($ \sim$ TeV) due to the (massive) KK modes, we have used the asymptotic form of the Bessel functions ($m_n \rightarrow 0$), and the fact that $e^{k \pi r_c} >> 1$ \cite{Contino:2002kc}.  If we consider $k >> m_{\Phi}$, we can use this form of (\ref{beautifulf}) in (\ref{closedsumofkkmodes}) to evaluate the integral in Eq.\ (\ref{PIQ2}). The correction to the coupling is given by:
\begin{eqnarray}
		\Delta_{\rm{1-loop}}^{\rm{scalar}}\left(q^2\right) = -\frac{C\left(r\right)}{6} \left\{\left(\alpha-1\right) \warp + \log\frac{\mu}{k}\right\}.
		\label{scalaroneloopwkkmodes}
\end{eqnarray}

From this analysis, we can construct the full form of the (energy dependent) SM couplings.  Luckily, this ``rather tedious" analysis has already been done \cite{Choi:2002ps}, and we will adapt these results to fit our purposes.  The arbitrary mass scale which was introduced in the regularization $\mu$ (\textit{cf} Eq.\ \ref{PIQ2}) becomes $M_*$---the scale of partial unification, and our cutoff.  Using Equation (\ref{choiestimate}), we find:
\begin{eqnarray}
	\nonumber
	\alpha_i^{-1}\left(p\right) &=& \frac{1}{\alpha_{\widetilde{SU}\left(5\right)}} + \mathcal{O}\left(\frac{1}{2 \pi}\right)
		\\
		\nonumber
		&&+~\frac{\frac{3}{2}}{12 \pi} \left\{-\warp + \log\frac{k}{M_*}\right\}
		\\
		\nonumber
		&&+~\frac{5}{24 \pi} \left\{22 \warp + 21 \log M_* \pi r_c\right\}
		\\
		\nonumber
		&&+~\frac{b_i}{2 \pi} \log \frac{M_*}{p} +\mathcal{O}\left(\frac{1}{2 \pi} \frac{M_*^2}{k^2}\right) \warp
		\\
		&&+~(\rm{5-d~threshold~effects}).
	\label{smrges}
\end{eqnarray}
The second line gives the contributions to the couplings from the massive scalar in the bulk---if we wish to modify the spectrum of the theory by adding additional higgses, we can add terms similar to these.  The third line gives the contributions from the $X$ and $Y$ gauge bosons of \fsu, and the last line is of the familiar form.  Note that the second and third lines give a universal (to each of the beta functions, independent of $p$) correction to $\alpha_{\widetilde{SU}\left(5\right)}^{-1}$, suggesting that we take
\begin{eqnarray}
	\alpha_{eff}^{-1} = \alpha_{\widetilde{SU}\left(5\right)}^{-1} + \Delta \alpha^{-1}.
	\label{effectivealpha}
\end{eqnarray}
The correction ($\Delta \alpha^{-1}$) is $\sim 45$ for the model presented here, and we will take $\alpha_{eff}^{-1} \sim 61$.\footnote{This value of $\alpha_{eff}^{-1}$ has been chosen because we want a value for the partial GUT coupling that is \textit{less than} the value of $\alpha_{U(1)}^{-1}\left(M_*\right)$, so we get unification of $SU(2) \times SU(3)$ before they unify with $U(1)$.  See Figure \ref{fig:betagraphs} below.}

We also notice the familiar SM runnings in the fourth line.  Because the fermions form complete representations in the GUT, they have no contributions to the runnings.  The SM gauge bosons, however, do contribute.  They are given as...
\begin{eqnarray}
\begin{array}{cccc}
	U(1):&b_1&=&0\\
	SU(2):&b_2&=&-\frac{22}{3}\\
	SU(3):&b_3&=&-11.
\end{array}
\end{eqnarray}
The $\mathcal{O}\left(\frac{1}{2 \pi} \frac{M_*^2}{k^2}\right) \warp$ terms are 5-d mass splittings that are calculable, but are of sub-sub-leading order. Thus, we do not calculate them here.

Finally, there are the 5-d threshold effects \cite{Agashe:2002pr} that are assumed to give the corrections needed for unification.  In order to break \fsu, we need the bulk scalar fields to take on vevs.  This is done by choosing a suitable potential for the fields, with minima at the desired mass scale.  It is this 5-d potential that gives the threshold corrections needed for unification. 

We see that the leading logarithm in each term of Equation (\ref{smrges}) is exactly as expected, from \cite{Pomarol:2000hp}.  Also notice that the terms proportional to $\warp$~are effects due to KK modes.  If we were to eliminate these states from the spectrum, we would recover the standard form of a coupling constant plus threshold corrections.\footnote{See, for example, \cite{RossGUTS}.}

Let us compare the runnings of the couplings in our model to those of the SM.  In the \fsu~models, $SU(5) \supset SU(3)_C \times SU(2)_L$.  We take the effective $SU(5)$ coupling, $\alpha_{eff}^{-1} = 61$.  The runnings of $SU(3)$ and $SU(2)$ are given by
\begin{subequations}
\label{runnings}
	\begin{eqnarray}
		\alpha_3^{-1}\left(E\right) = \alpha_{eff}^{-1} + \frac{b_3}{2 \pi} \log\frac{M_*}{E} + \frac{\delta b_3}{2 \pi} \log\frac{M_*}{E},\\
		\alpha_2^{-1}\left(E\right) = \alpha_{eff}^{-1} + \frac{b_2}{2 \pi} \log\frac{M_*}{E} + \frac{\delta b_2}{2 \pi} \log\frac{M_*}{E}.
	\end{eqnarray}
\end{subequations}
We can calculate the $\delta b_i$'s needed for unification using $\alpha_3\left(M_Z\right) \cong 0.1187\pm0.0020$, $\alpha_2\left(M_Z\right) \cong 0.033961 \pm 0.000006$, and $\alpha_1\cong0.017022 \pm 0.000002$ \cite{pdg}. From this we find that
\begin{subequations}
\begin{eqnarray}
	SU(3):~\delta b_3&\cong&0.761,\\
	SU(2):~\delta b_2&\cong&0.185.	
\end{eqnarray}
\end{subequations}
These expressions are plotted in Figure \ref{fig:betagraphs}, using $k \sim 10^{18}$ GeV and $M_* \sim 3 \times 10^{15}$ GeV.  We have added threshold effects on the order of about 10\% to the $U(1)_Y$ in the SM, because we expect that the corrections to the coupling are of the same size as those of the other graphs.

\begin{figure}[!t]
  \centering
  	\subfigure[]{\includegraphics[scale=0.13]{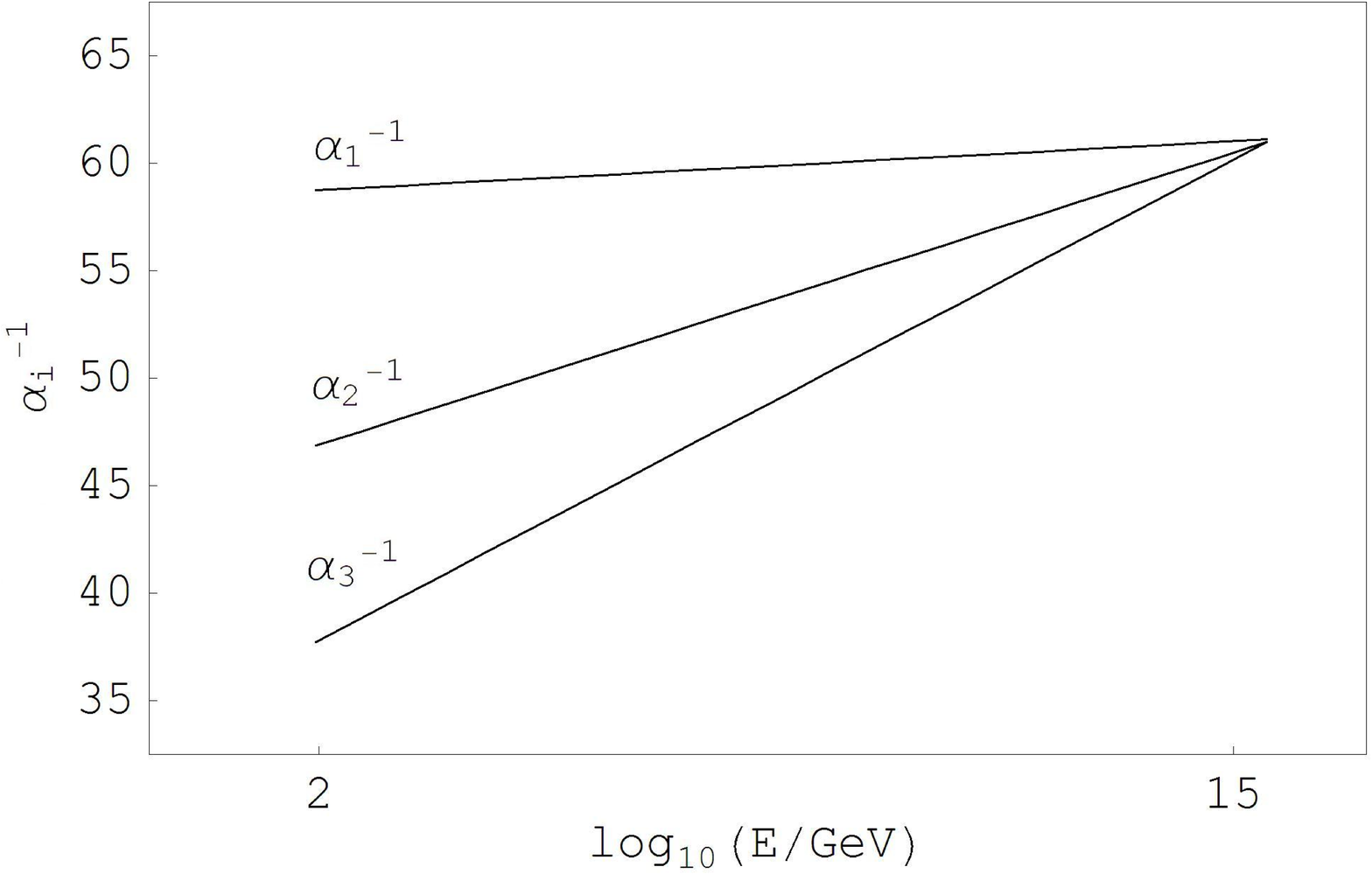}}         
  	\subfigure[]{\includegraphics[scale=0.13]{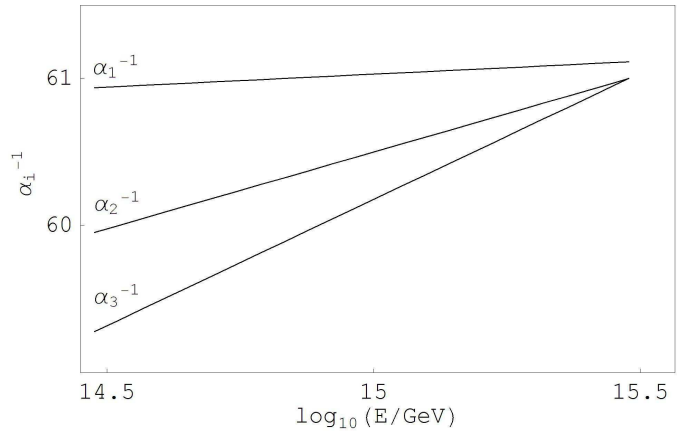}}
  \caption[SM beta functions in our model.]{SM beta functions in our model, with corrections from KK modes, GUT scalars, and GUT-mass bosons.  Note that the unification at $M_* \sim 3\times 10^{15}$ GeV is not exact---only $SU(3)$ and $SU(2)$ unify here, as expected.}
  \label{fig:betagraphs}
\end{figure}
\subsection{Constraining Randall-Sundrum GUTs}
\label{subsec:constraints}

We will now briefly comment on an interesting constraint revealed in our investigations of this model.  It is hoped that the constraints shown here will be of use to anyone wishing to construct RS1 GUTs.  Further, while it was not explicitly investigated, we believe that these constraints apply also to more general 5-d orbifold GUTs, as well as any construction that gives universal corrections to the beta functions of the model.  In general, the contributions to the beta functions coming from the KK modes of the bulk scalars and the GUT gauge bosons are:
\beqn
	\frac{\Delta_{\rm{1-loop}}}{2 \pi} \equiv \frac{C(r)}{12 \pi} \left\{-\warp + \log\frac{k}{M_*}\right\} +\frac{C(N)}{24 \pi} \left\{22 \warp + 21 \log M_* \pi r_c\right\}.
\eeqn
$C(r)$ is the Dynkin index of the scalar field, and $C(N)$ is the quadratic Casimir operator of the algebra.  In what follows, we will look exclusively at $SU(5)$, where $C(N) = 5$.

In some sense, the threshold effects due to the presence of a bulk scalar ``compete" with the corrections due to the $SU(5)$ gauge bosons---they give contributions of opposite sign.  The scalar loops are weighted by their Dynkin index, a group theory factor that depends on the representation in which the scalars transform.  One can compute these, or just look them up \cite{Slansky:1981yr}, and find that larger representations (generally) have larger Dynkin indices.  

\begin{table}
	\centering
	\begin{tabular}{cccccc}
		\hline
		\hline
		Model&Bulk Scalars&$\sum_r C(r)$&$M_{GUT}$ (GeV)&$k$ (GeV)&$\Delta / 2 \pi$\\
		\hline
		RS1 $\widetilde{SU}(5)$&\textbf{10}&1.5&$3 \times 10^{15}$&$10^{18}$&44.796\\
		Georgi \cite{Georgi:1974sy}&\textbf{24}&5&$3 \times 10^{15}$&$10^{18}$&38.956\\
		Dorsner \cite{Dorsner:2006zq, DorsnerAAA, DorsnerBBB}&\textbf{15},\textbf{24}&8.5&$3 \times 10^{15}$&$10^{18}$&33.143\\
		HHM I \cite{Hayashi:1982mn}&\textbf{24}, \textbf{45}&17&$3 \times 10^{15}$&$10^{18}$&19.026\\
		HHM II \cite{Hayashi:1982mn}&\textbf{45}, \textbf{75}&37&$3 \times 10^{15}$&$10^{18}$&-14.191\\		
		\hline
	\end{tabular}
	\caption[The scalar content of various $SU(5)$ models.]{We look at the scalar content of various non-supersymmetric $SU(5)$ constructions, to get some idea of the representations that are important for model-building.  As we include larger and larger bulk scalar reps, the universal contributions to the beta functions approach zero.  Calculated here are the contributions from scalars in the adjoint (\textbf{24}) and other various reps.}
	\label{tab:one}
\end{table}

The bulk field content of the model will govern the types of values that one can obtain for $M_*$ and $\alpha_{\widetilde{SU}\left(5\right)}^{-1}$, via universal contributions to the beta functions from the bulk matter, as in Equation (\ref{effectivealpha}).  Likewise, any constraints on $\alpha_{\widetilde{SU}\left(5\right)}^{-1}$ will tell us the maximum contributions from KK threshold effects, as per \cite{Choi:2002ps}.  Any effects from some higher unification scale, at $M_{GUT}$, would still enter the beta function as a correction to the effective $\alpha_{eff}^{-1}$, and be of the same form as Equation (\ref{smrges}).  This will also put constraints on the size and number of bulk fields introduced in the GUT model.  In Table \ref{tab:one} we have looked at the scalar content of some 4-d $SU(5)$ theories.  We note that these models were not built within the RS1 framework, but we have looked at these examples to get an idea for the important scalar reps used in model building.

In the model presented in this paper, the corrections due to scalars and vectors are $\sim 45$.  Requiring that our threshold effects be on the order of 10\% means that $55 \lesssim \alpha_{\widetilde{SU}\left(5\right)}^{-1} \lesssim 62$, which in turn forces $10 \lesssim \alpha_{\widetilde{SU}\left(5\right)}^{-1} \lesssim 17$.  A more interesting case is when the corrections due to the KK modes are negative, as in the HHM II model \cite{Hayashi:1982mn}. This  occurs as we include larger (or more) scalar reps in our models.  Then, the value of $\alpha_{SU(5)}^{-1}$ must be at least as large as the corrections coming from the bulk scalar and vector representations in order to ensure the positivity of $\alpha_{eff}^{-1}$, as per Equation (\ref{effectivealpha}).  If we were to build an $SU(5)$ RS1 GUT, placing the matter content of HHM II \cite{Hayashi:1982mn} in the bulk, we could plot Equation (\ref{effectivealpha}), showing where $\alpha_{SU(5)}^{-1}$ becomes negative---see Figure \ref{fig:alphaeff}.  This gives the possibility of excluding this model, based on estimates of the size of $\alpha_5^{-1}$, or equivalently $g_5$ from Equation (\ref{choiestimate}).
  
\begin{figure}[t]
	\centering
	\includegraphics[width=15cm]{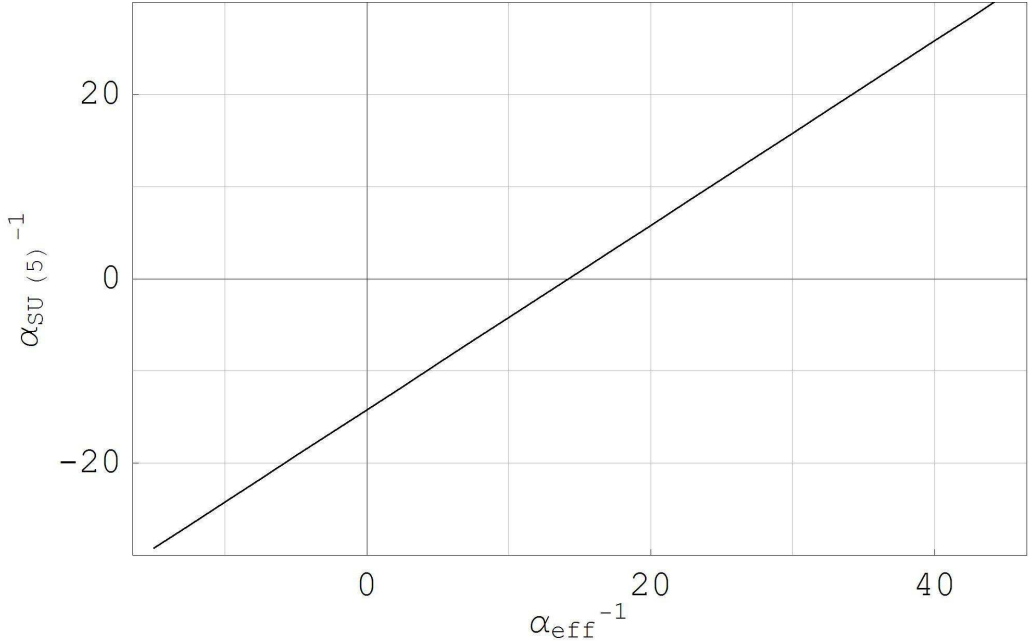}
	\caption[A plot of Equation (\ref{effectivealpha}).]{A plot of Equation (\ref{effectivealpha}) for the model HHM II \cite{Hayashi:1982mn}.  If we choose $\alpha_{eff}^{-1}$ less than 15, we get a negative value for $\alpha_{SU\left(5\right)}^{-1}$, which is unphysical.}
	\label{fig:alphaeff}
\end{figure}

\subsection{Proton Lifetime}
\label{subsec:numassandpdecay}
Without some exponential tuning of Yukawa couplings, proton decay will be a problem as per Equation (\ref{bdopinleeft}).  When solving the hierarchy in the higgs sector of the model by introducing the warp factor, we must ask ourselves if we are willing to introduce another fine-tuning in the form of an exponentially small Yukawa coupling.  Ideally we would like to explain all things in terms of $\mathcal{O}(1)$ parameters, and if we insist on using bulk scalar fields to break \fsu~, we may have to take a more creative approach, invoking some (possibly discrete) symmetry to protect baryons in the low energy effective field theory by forbidding terms like Equation (\ref{bdopinleeft}).  If proton decays are observed in a next generation experiment, we will almost certainly have to accept some fine tuning.  

Further, in the RS1 scenario, one must be careful to check \textit{all} of the possible decay modes of the proton---there will be new decays through KK mode exchange.  In general, the KK modes will have masses on the order of a few TeV.  The SM fermions may interact with these KK modes to violate bounds on proton decay, and may even produce flavor changing neutral currents at an unacceptable rate \cite{Agashe:2004bm}.  The problem is not limited to RS1 GUTs, but also to the RS1 formulation of the SM \cite{Gherghetta:2000qt}.  The only way to eliminate these problems is to break the GUT symmetry with boundary conditions \cite{Goldberger:2002pc, Agashe:2002pr}, or invoke some discrete or global symmetry which protects baryon and lepton number.

In the standard \fsu~models, the predominant baryon decay operator is given by \cite{Ellis:1988tx}:
\beqn
	\mathcal{L}\sim\frac{g_{\widetilde{SU}\left(5\right)}^2}{M_*^2} \left\{-\bar{d} \gamma^{\mu} d \bar{u} \gamma_{\mu} \nu + \bar{d} \gamma^{\mu} u \bar{u} \gamma_{\mu} \ell^-\right\},
	\label{fsu5protondecay}
\eeqn
where $\ell^-$ is a linear combination of the three (left-handed) leptons.  The typical calculation, using $\ell \sim e$, puts the \fsu\ prediction for the proton lifetime
\beqn
	\tau_{p\rightarrow e^+ \pi^0} \sim 10^{33-37} \rm{~y},
\eeqn
which safely evades the current lower bounds.  Let us quickly estimate the proton lifetime using our choice of constants, that is, $\alpha_{eff}^{-1} \sim 61$ and $M_* \sim 3 \times 10^{15}$ GeV.  Ellis, Lopez and Nanopoulos \cite{Ellis:1995at} have estimated the lifetime of the proton in \fsu~via the decay channel $p\rightarrow e^+ \pi^0$:
\beqn
	\tau_{p\rightarrow e^+ \pi^0} \cong 1.5 \times 10^{33} \left(\frac{M_*}{10^{15}\rm{~GeV}}\right)^4 \left(\frac{0.042}{\alpha_{\widetilde{SU}\left(5\right)}}\right)^2 \rm{~y}.
\eeqn
For our choice of values for $M_*$ and $\alpha_{eff}^{-1}$, we find
\beqn
	\tau_{p\rightarrow e^+ \pi^0} \sim 8.0 \times 10^{35} \rm{~y}.
\eeqn

Now, Equation (\ref{fsu5protondecay}) and Equation (\ref{bdopinleeft}) give the same term in the effective lagrangian.  We may write:
\beqn
	\mathcal{L}_{\Delta B \neq 0} \sim \left(\frac{2k}{M_5^3}\frac{\lambda}{N_i N_j N_k N_l} \frac{e^{\delta\warp}}{\delta}\right) \bar{\psi}_{\mathbf{\bar{5}}} \psi_{\mathbf{5}} \bar{\psi}_{\mathbf{\overline{10}}} \psi_{\mathbf{10}},
\eeqn
where $\delta > 1$ in general, and $\lambda$ is some (dimensionless) coupling in the fundamental theory.  Comparing with Equation (\ref{fsu5protondecay}), with the replacement $M_{Pl}\rightarrow M_*$, we have\footnote{We will leave out factors of $\mathcal{O}(1)$, for comparison's sake.}
\beqn
	g_{\widetilde{SU}\left(5\right)}^2\sim \frac{\lambda}{N_i N_j N_k N_l}\frac{e^{\delta\warp}}{\delta}.
\eeqn
Because the value of $\delta$ must be calculated from the UV completion of the theory, we cannot determine an accurate value for the proton lifetime.  We know, however, that $\delta$ is $\mathcal{O}(\frac{1}{2})$, and using the relation between 4-d and 5-d mass scales, $M_5$ is found to be within two or three orders of magnitude of $M_*$. In combination with the parameters from Section \ref{subsec:gcr}, this requires that
\beqn
	\lambda \sim 10^{-44},
\eeqn
in order to match the predictions of the \fsu~GUTs.  This is why we were so cavalier with two or three orders of magnitude and a few numbers of order 1!  We have created a fine tuning problem thirty orders of magnitude worse than the QCD $CP$ problem.  If there exists a (D=6) baryon decay term like Equation (\ref{bdopinleeft}) in the GUT, then the Yukawa coupling must be tremendously small to keep the proton sufficiently long lived.  Thus, we most likely must invoke discrete symmetries to remove such finely tuned terms, or rely on some other mechanism to eliminate proton decay.  Specifically, it has been pointed out \cite{Dorsner:2004jj} that non-diagonal Yukawa couplings can lead to a stable proton---in effect, proton decay is ``rotated'' away.\footnote{We would like to thank I. Dorsner for pointig this out.}


\section{\label{sec:conclusions}Conclusions and outlook}
We have presented a \fsu~model within the framework of the RS1 proposal, which retains much of the successes of the standard \fsu~phenomenology.  The higgs content is minimal, utilizing only the \textbf{10} rep to break the GUT symmetry.  The fermions lie in the normal three generations of \fsu~reps, with no extra copies.

One future direction is the possible embedding of our model into some larger symmetry, like $SO(10)$ or $E_6$.  Indeed, the seminal \fsu~work \cite{Barr:1981qv, Derendinger:1983aj} was done as an alternative breaking pattern for $SO(10)$.  There are also proposals for $\widetilde{SO}(10)$ \cite{Tamvakis:1987sd,Kataoka:1992th,Sato:1993ir,Shimojo:1994qk,Maekawa:2003wm,Huang:2006nu} and $\widetilde{E}_6$ \cite{Eeg:1989gv,Eeg:1989ev,Das:2006ib} (flipped $SO(10)$ and flipped $E_6$, respectively) .  

Finally, the proliferation of representations of non-flipped RS1 GUT models, discussed in Section \ref{sec:breakingsu5}, may be alleviated through models in which $E_6 \rightarrow SO(10) \times U(1)_X$, with $U(1)_X$ plays the role of a ``family" symmetry. This $U(1)$ may be reinterpretted so as to provide the copies of the representations needed in, for example, \cite{Agashe:2004bm}. In $E_6$, the \textbf{27} contains three copies of the \textbf{16} in $SO(10)$.  Finally, a supersymmeterized version of this model may be attempted.  If so, some symmetry protecting $B-L$ should be found or imposed. (Alternately, this model could be reformulated using boundary conditions to break the GUT symmetry.)

\begin{acknowledgments}
The authors would like to thank Tibra Ali and Anzhong Wang for helpful conversations.
\end{acknowledgments}



\begin{thebibliography}{10}

\bibitem{Randall:1999ee}
L.~Randall and R.~Sundrum,
\newblock Phys. Rev. Lett. {\bf 83}, 3370 (1999).

\bibitem{Goldberger:1999uk}
W.~D. Goldberger and M.~B. Wise,
\newblock Phys. Rev. Lett. {\bf 83}, 4922 (1999).

\bibitem{Grossman:1999ra}
Y.~Grossman and M.~Neubert,
\newblock Phys. Lett. {\bf B474}, 361 (2000).

\bibitem{Davoudiasl:1999tf}
H.~Davoudiasl, J.~L. Hewett, and T.~G. Rizzo,
\newblock Phys. Lett. {\bf B473}, 43 (2000).

\bibitem{Pomarol:1999ad}
A.~Pomarol,
\newblock Phys. Lett. {\bf B486}, 153 (2000).

\bibitem{Georgi:1974sy}
H.~Georgi and S.~L. Glashow,
\newblock Phys. Rev. Lett. {\bf 32}, 438 (1974).

\bibitem{Mohapatra:1986uf}
R.~N. Mohapatra,
\newblock Berlin, Germany: Springer (1986) 309. (Contemporary Physics).

\bibitem{Agashe:2002pr}
K.~Agashe, A.~Delgado, and R.~Sundrum,
\newblock Ann. Phys. {\bf 304}, 145 (2003).

\bibitem{Goldberger:2002pc}
W.~D. Goldberger, Y.~Nomura, and D.~R. Smith,
\newblock Phys. Rev. {\bf D67}, 075021 (2003).

\bibitem{Agashe:2004bm}
K.~Agashe and G.~Servant,
\newblock JCAP {\bf 0502}, 002 (2005).

\bibitem{Hayato:1999az}
Y.~Hayato et~al.,
\newblock Phys. Rev. Lett. {\bf 83}, 1529 (1999).

\bibitem{Contino:2002kc}
R.~Contino, P.~Creminelli, and E.~Trincherini,
\newblock JHEP {\bf 10}, 029 (2002).

\bibitem{Barr:1981qv}
S.~M. Barr,
\newblock Phys. Lett. {\bf B112}, 219 (1982).

\bibitem{Derendinger:1983aj}
J.~P. Derendinger, J.~E. Kim, and D.~V. Nanopoulos,
\newblock Phys. Lett. {\bf B139}, 170 (1984).

\bibitem{Antoniadis:1987tv}
I.~Antoniadis, J.~R. Ellis, J.~S. Hagelin, and D.~V. Nanopoulos,
\newblock Phys. Lett. {\bf B205}, 459 (1988).

\bibitem{Antoniadis:1988tt}
I.~Antoniadis, J.~R. Ellis, J.~S. Hagelin, and D.~V. Nanopoulos,
\newblock Phys. Lett. {\bf B208}, 209 (1988).

\bibitem{Antoniadis:1989zy}
I.~Antoniadis, J.~R. Ellis, J.~S. Hagelin, and D.~V. Nanopoulos,
\newblock Phys. Lett. {\bf B231}, 65 (1989).

\bibitem{Leontaris:1990bw}
G.~K. Leontaris, J.~Rizos, and K.~Tamvakis,
\newblock Phys. Lett. {\bf B243}, 220 (1990).

\bibitem{Bailin:1990qb}
D.~Bailin, E.~K. Katechou, and A.~Love,
\newblock Int. J. Mod. Phys. {\bf A7}, 153 (1992).

\bibitem{Burwick:1990vy}
T.~T. Burwick, R.~K. Kaiser, and H.~F. Muller,
\newblock Nucl. Phys. {\bf B362}, 232 (1991).

\bibitem{Rizos:1990xn}
J.~Rizos and K.~Tamvakis,
\newblock Phys. Lett. {\bf B251}, 369 (1990).

\bibitem{Lopez:1991ac}
J.~L. Lopez and D.~V. Nanopoulos,
\newblock Phys. Lett. {\bf B268}, 359 (1991).

\bibitem{Lopez:1991ec}
  J.~L.~Lopez and D.~V.~Nanopoulos,
  \newblock arXiv:hep-th/9110036.

\bibitem{Antoniadis:1991fc}
I.~Antoniadis, J.~Rizos, and K.~Tamvakis,
\newblock Phys. Lett. {\bf B278}, 257 (1992).

\bibitem{Lopez:1992kg}
J.~L. Lopez, D.~V. Nanopoulos, and K.-j. Yuan,
\newblock Nucl. Phys. {\bf B399}, 654 (1993).

\bibitem{Hatzinikitas:1992ip}
A.~Hatzinikitas and A.~Toon,
\newblock Int. J. Mod. Phys. {\bf A8}, 557 (1993).

\bibitem{Lopez:1995pr}
  J.~L.~Lopez and D.~V.~Nanopoulos,
  \newblock arXiv:hep-ph/9511266.
  
\bibitem{Lopez:1997hq}
  J.~L.~Lopez and D.~V.~Nanopoulos,
  \newblock arXiv:hep-ph/9701264.
  
\bibitem{Ellis:1997ni}
J.~R. Ellis, G.~K. Leontaris, S.~Lola, and D.~V. Nanopoulos,
\newblock Phys. Lett. {\bf B425}, 86 (1998).

\bibitem{Ellis:1999ce}
J.~R. Ellis, G.~K. Leontaris, and J.~Rizos,
\newblock Phys. Lett. {\bf B464}, 62 (1999).

\bibitem{Cleaver:2000sc}
G.~B. Cleaver, J.~R. Ellis, and D.~V. Nanopoulos,
\newblock Nucl. Phys. {\bf B600}, 315 (2001).

\bibitem{Faraggi:2002ah}
A.~E. Faraggi, R.~Garavuso, and J.~M. Isidro,
\newblock Nucl. Phys. {\bf B641}, 111 (2002).

\bibitem{Faraggi:2002dg}
  A.~E.~Faraggi, R.~Garavuso and J.~M.~Isidro,
  \newblock arXiv:hep-th/0209245.
  
\bibitem{Chen:2005ab}
C.~M. Chen, G.~V. Kraniotis, V.~E. Mayes, D.~V. Nanopoulos, and J.~W. Walker,
\newblock Phys. Lett. {\bf B611}, 156 (2005).

\bibitem{Ellis:2004cj}
J.~R. Ellis, V.~E. Mayes, and D.~V. Nanopoulos,
\newblock Phys. Rev. {\bf D70}, 075015 (2004).

\bibitem{Chen:2005mm}
C.~M. Chen, G.~V. Kraniotis, V.~E. Mayes, D.~V. Nanopoulos, and J.~W. Walker,
\newblock Phys. Lett. {\bf B625}, 96 (2005).

\bibitem{Chen:2005cf}
C.-M. Chen, V.~E. Mayes, and D.~V. Nanopoulos,
\newblock Phys. Lett. {\bf B633}, 618 (2006).

\bibitem{Chen:2006ip}
C.-M. Chen, T.~Li, and D.~V. Nanopoulos,
\newblock Nucl. Phys. {\bf B751}, 260 (2006).

\bibitem{Huang:2006nu}
  C.~S.~Huang, T.~Li, C.~Liu, J.~P.~Shock, F.~Wu and Y.~L.~Wu,
\newblock  JHEP {\bf 0610}, 035 (2006)

\bibitem{Shafi:2006dm}
  Q.~Shafi and Z.~Tavartkiladze,
  \newblock arXiv:hep-ph/0606188.
  
\bibitem{Cvetic:2006by}
  M.~Cvetic and P.~Langacker,
  \newblock arXiv:hep-th/0607238.

\bibitem{Kim:2006hw}
  J.~E.~Kim and B.~Kyae,
  \newblock arXiv:hep-th/0608086.

\bibitem{Kim:2006hv}
  J.~E.~Kim and B.~Kyae,
  \newblock arXiv:hep-th/0608085.

\bibitem{Blumenhagen:2006ux}
  R.~Blumenhagen, S.~Moster and T.~Weigand,
  \newblock Nucl.\ Phys.\ B {\bf 751}, 186 (2006).

\bibitem{Cleaver:1998sa}
G.~B. Cleaver, A.~E. Faraggi, and D.~V. Nanopoulos,
\newblock Phys. Lett. {\bf B455}, 135 (1999).

\bibitem{Kawai:1986ah}
H.~Kawai, D.~C. Lewellen, and S.~H.~H. Tye,
\newblock Nucl. Phys. {\bf B288}, 1 (1987).

\bibitem{Antoniadis:1986rn}
I.~Antoniadis, C.~P. Bachas, and C.~Kounnas,
\newblock Nucl. Phys. {\bf B289}, 87 (1987).

\bibitem{Kawai:1987ew}
H.~Kawai, D.~C. Lewellen, J.~A. Schwartz, and S.~H.~H. Tye,
\newblock Nucl. Phys. {\bf B299}, 431 (1988).

\bibitem{Antoniadis:1987wp}
I.~Antoniadis and C.~Bachas,
\newblock Nucl. Phys. {\bf B298}, 586 (1988).

\bibitem{Barr:2002fb}
S.~M. Barr and I.~Dorsner,
\newblock Phys. Rev. {\bf D66}, 065013 (2002).

\bibitem{Dorsner:2003yg}
I.~Dorsner,
\newblock Phys. Rev. {\bf D69}, 056003 (2004).

\bibitem{Agashe:2002bx}
K.~Agashe, A.~Delgado, and R.~Sundrum,
\newblock Nucl. Phys. {\bf B643}, 172 (2002).

\bibitem{Choi:2002ps}
K.-w. Choi and I.-W. Kim,
\newblock Phys. Rev. {\bf D67}, 045005 (2003).

\bibitem{Goldberger:2002hb}
W.~D. Goldberger and I.~Z. Rothstein,
\newblock Phys. Rev. {\bf D68}, 125011 (2003).

\bibitem{Gherghetta:2000qt}
T.~Gherghetta and A.~Pomarol,
\newblock Nucl. Phys. {\bf B586}, 141 (2000).

\bibitem{jaxodraw}
D.~Binosi and L.~Theu\ss{}l,
\newblock Comp. Phys. Comm. {\bf 161}, 76 (2004).

\bibitem{Pomarol:2000hp}
A.~Pomarol,
\newblock Phys. Rev. Lett. {\bf 85}, 4004 (2000).

\bibitem{RossGUTS}
G.~G. Ross,
\newblock {\em Grand Unified Theories},
\newblock Westview Press, Oxford, UK, 1984.

\bibitem{pdg}
S.~Eidelman et~al.,
\newblock Phys. Lett. {\bf B592}, 1 (2004).

\bibitem{Slansky:1981yr}
R.~Slansky,
\newblock Phys. Rept. {\bf 79}, 1 (1981).

\bibitem{Dorsner:2006zq}
I.~Dorsner,
\newblock Nucl. Phys. {\bf B723}, 53 (2005).

\bibitem{DorsnerAAA}
I.~Dorsner,
\newblock Nucl. Phys. {\bf B747}, 312 (2006).

\bibitem{DorsnerBBB}
I.~Dorsner,
\newblock hep-ph/0606062.

\bibitem{Hayashi:1982mn}
H.~Hayashi, A.~Murayama, and M.~J. Hayashi,
\newblock Phys. Rev. {\bf D26}, 1185 (1982).

\bibitem{Ellis:1988tx}
J.~R. Ellis, J.~S. Hagelin, S.~Kelley, and D.~V. Nanopoulos,
\newblock Nucl. Phys. {\bf B311}, 1 (1988).

\bibitem{Ellis:1995at}
J.~R. Ellis, J.~L. Lopez, and D.~V. Nanopoulos,
\newblock Phys. Lett. {\bf B371}, 65 (1996).

\bibitem{Tamvakis:1987sd}
K.~Tamvakis,
\newblock Phys. Lett. {\bf B201}, 95 (1988).

\bibitem{Kataoka:1992th}
H.~Kataoka, H.~Munakata, H.~Sato, and S.~Tanaka,
\newblock Phys. Lett. {\bf B289}, 321 (1992).

\bibitem{Sato:1993ir}
H.~Sato and M.~Shimojo,
\newblock Phys. Rev. {\bf D48}, 5798 (1993).

\bibitem{Shimojo:1994qk}
M.~Shimojo,
\newblock Prog. Theor. Phys. {\bf 92}, 413 (1994).

\bibitem{Maekawa:2003wm}
N.~Maekawa and T.~Yamashita,
\newblock Phys. Lett. {\bf B567}, 330 (2003).

\bibitem{Eeg:1989gv}
J.~O. Eeg,
\newblock Z. Phys. {\bf C46}, 665 (1990).

\bibitem{Eeg:1989ev}
J.~O. Eeg,
\newblock Fizika {\bf 21}, 22 (1989).

\bibitem{Das:2006ib}
C.~R. Das and L.~V. Laperashvili,
\newblock (2006).

\bibitem{Dorsner:2004jj}
  I.~Dorsner and P.~Fileviez Perez,
  \newblock Phys.\ Lett.\ {\bf B606}, 367 (2005).
   
\end{thebibliography}

%
%
%
%
%
%
\end{document}